\documentclass[12pt]{article}

\usepackage{fullpage}
\usepackage{times}
\usepackage{amsmath,amsthm}
\usepackage{epsfig}

\newsavebox{\fmbox}
\newsavebox{\algobox}

\newtheorem{theorem}{Theorem}
\newtheorem{definition}{Definition}
\newtheorem{corollary}{Corollary}
\newtheorem{lemma}{Lemma}

\renewcommand{\epsilon}{\varepsilon}
\newcommand{\A}{{\cal A}}
\newcommand{\B}{{\cal B}}

\newcommand{\C}{{\rm C}}

\newcommand{\rw}{c}
\newcommand{\cl}{r}
\newcommand{\brw}{{\bf c}}
\newcommand{\bcl}{{\bf r}}
\newcommand{\gig}{G}
\newcommand{\vc}{a}

\newcommand{\vvv}{{\bf z}}
\newcommand{\lb}{\omega}

\newcommand{\rf}{{\rm ref}}

\newcommand{\RW}{C}
\newcommand{\CL}{R}

\newcommand{\amp}{{\rm amp}} 

\begin{document}

\title{Spectra of Quantized Walks and a $\sqrt{\delta\epsilon}$-Rule}

\author{
Mario Szegedy\thanks{
Rutgers University,
email: {\tt szegedy@cs.rutgers.edu}; 
This work was supported by NSF grant 0105692,
and in part by the National
Security Agency (NSA) and Advanced Research and Development Activity
(ARDA)
under Army Research Office (ARO) contract number DAAD19-01-1-0506.
}}
\date{}

\maketitle

\begin{abstract}
We introduce quantized bipartite walks,
compute their spectra,
generalize the algorithms of Grover \cite{g} and Ambainis \cite{amb03}
and interpret them as
quantum walks with memory.
We compare the performance of walk based 
classical and quantum algorithms and show that the latter run
much quicker in general. 
Let $P$ be a symmetric Markov chain with transition probabilities
$P[i,j]$, $(i ,j\in [n])$. Some elements
of the state space 
are marked. We are promised that the set of marked
elements has size either zero or at least $\epsilon n$. The
goal is to find out with great certainty which of the above two cases
holds. Our model is a black box
that can answer certain yes/no questions and can generate random elements
picked from certain distributions. 
More specifically, 
by request the black box can give us a 
uniformly distributed
random element for the cost of $\wp_{0}$.
Also, when ``inserting'' an element $i$
into the black box we can obtain a random element 
$j$, where $j$ is distributed 
according to $P[i,j]$. The cost of the latter 
operation is $\wp_{1}$.
Finally, we can use the black box to test 
if an element $i$ is marked,
and this costs us $\wp_{2}$.
If $\delta$ is the eigenvalue gap of $P$,
there is a simple classical algorithm with cost
$O(\wp_{0} + (\wp_{1}+\wp_{2})/\delta\epsilon)$
that solves the above promise problem.
(The algorithm is efficient if $\wp_{0}$
is much larger than $\wp_{1}+\wp_{2}$.) In contrast,
we show that for the ``quantized'' version of the algorithm 
it costs only
$O(\wp_{0} + (\wp_{1}+\wp_{2})/\sqrt{\delta\epsilon})$ 
to solve the problem.
We refer to this as the 
$\sqrt{\delta\epsilon}$
rule. Among the technical contributions we give a formula 
for the spectrum of the product of two general reflections.
\end{abstract}

\section{Introduction}

The recent algorithm of Ambainis \cite{amb03} 
which exploits properties of quantum walks on versions of
Johnson graphs
opens up a new avenue for walk based algorithm designs.
It claims as its direct predecessor the algorithm
of Grover \cite{g} for database search.
Ambainis's algorithm, like \cite{g}, is a search algorithm,
but for many problems to which it applies, such as element distinctness,
it runs quicker than any application
of \cite{g}.
In the present article we give a thorough analysis of the type of walks
that power the construction in \cite{amb03},
and use them in a very general algorithmic scheme.
In our efforts to simplify the proofs
we modify the paradigm that Grover has set up.

Ambainis's idea that a quantum walk can accelerate
black box search for pairs of colliding elements
demonstrates that
diffusion-based quantum walks are more versatile
than it had been thought previously. 
The consequences are numerous. Based on it
Magniez, Santha and Szegedy \cite{mss} give a quicker algorithm
for the oracle version of the triangle finding problem.
Childs and Eisenberg \cite{ce}
streamline the analysis of 
\cite{amb03}, and give further examples to its use. 
These papers, even the latter one 
do not change the proof structure of \cite{amb03}
significantly.

In contrast, our discussion departs 
from that of \cite{amb03} at several points:
1. We treat not only the Johnson graphs but all Markov chains;
2. We show how to eliminate the restriction
Ambainis imposes on the state of the machine: we no longer need to
request that it stays in a constant dimensional subspace
of the entire state space.
3. We circumvent relevant parts of the proof by introducing
``memory'' for walks. 
As a consequence of 1., 2., and 3.
we obtain:

Let $P$ be a symmetric Markov chain with transition probabilities
$P[i,j]$, $(i ,j\in [n])$. Some elements
of the state space 
are marked. We are promised that the set of marked
elements has size either zero or at least $\epsilon n$. The
goal is to find out with great certainty which of the above two cases
holds. Our model is a black box
that can answer certain yes/no questions and can generate random elements
picked from certain distributions. 
More specifically, 
by request the black box can give us a 
uniformly distributed
random element for the cost of $\wp_{0}$.
Also, when ``inserting'' an element $i$
into the black box we can obtain a random element 
$j$, where $j$ is distributed 
according to $P[i,j]$. The cost of the latter 
operation is $\wp_{1}$.
Finally, we can use the black box to test 
if an element $i$ is marked,
and this costs us $\wp_{2}$.
If $\delta$ is the eigenvalue gap of $P$,
there is a simple classical algorithm with cost
$O(\wp_{0} + (\wp_{1}+\wp_{2})/\delta\epsilon)$
that solves the above promise problem.
(The algorithm is efficient if $\wp_{0}$
is much larger than $\wp_{1}+\wp_{2}$.) In contrast,
we show that for the ``quantized'' version of the algorithm 
it costs only
$O(\wp_{0} + (\wp_{1}+\wp_{2})/\sqrt{\delta\epsilon})$ 
to solve the problem.
We refer to this as the 
$\sqrt{\delta\epsilon}$
rule. Among the technical contributions we give a formula 
for the spectrum of the product of two general reflections.

We settle at a
walk model, directly derived from \cite{amb03}, 
that we call ``bipartite.''
We focus only on walks that do diffusion for coin flip.
Our formulas for the spectra and the eigenvalues of these walks
should be useful in other contexts as well.

In the quantum walk literature one
can find two separate directions. {\em Discrete} time walks were introduced by
Y. Aharonov, L. Davidovich, and N. Zagury \cite{adz} and
re-introduced by D. A. Meyer \cite{mey}.
The properties of these walks were studied in one dimension by
Ambainis, E. Bach, A. Nayak, A. Vishwanath, and J. Watrous: 
\cite{abnv}, and in general by 
D. Aharonov, A. Ambainis, J. Kempe, U. Vazirani \cite{aakv}.

{\em Continuous} time walks were introduced by
E. Farhi, S. Gutmann \cite{fg}, and they were studied
by J. Roland, N. Cerf \cite{rc}, by
Wim van Dam, Michele Mosca, Umesh V. Vazirani \cite{dmv},
by Childs and Goldstone \cite{cg} and many others.
Andrew M. Childs, Richard Cleve, Enrico Deotto, 
Edward Farhi, Sam Gutmann, Daniel A. Spielman \cite{ccdf}
show an example where a continuous quantum walk exponentially 
quicker traverses a graph than its deterministic counterpart.
Continuous walks are intimately
related to the paradigm of adiabatic computation.

In this article we are concerned about special discrete time walks.
The results are self-contained.

\section{Notations}

We develop notations for bipartite objects,
both classical and quantum. The symbol $[n|$ denotes a 
{\em left}-set of size $n$ and symbol $|m]$ denotes a
{\em right}-set of size $m$. A classical bipartite walk
takes place on the node set $[n|\cup|m]$:

\medskip

\begin{center}
\begin{tabular}{l|ccc}
 &  Left & \hspace{0.5in} & Right \\\hline
 & & & \\
Set: & $[n|$ & & $|m]$ \\
Vector: &  $a = (\alpha_{1},\ldots,\alpha_{n}|$ & & 
$b=|\beta_{1},\ldots,\beta_{m})$.
\end{tabular}
\end{center}

\medskip

\noindent 
We can concatenate left and right vectors into vectors of length $n+m$:
\[
(a,b) = (\alpha_{1},\ldots,\alpha_{n},\beta_{1},\ldots,\beta_{m}).
\]
Left and right vectors and their concatenations 
are classical objects. To quantize them we
form tensor products from them that lie in the 
Hilbert space ${\bf C}^{[n|\times |m]} = {\bf C}^{[n|}\otimes {\bf C}^{|m]}$.

Among the other not entirely standard notations we need are $\circ$,
which means point-wise product of vectors and matrices and
the square root of a vector or a matrix, which means point-wise
square root:
\[
\sqrt{M}[i,j] = \sqrt{M[i,j]}.
\] 
We apply the latter operation only if the elements of the
matrix or vector are non-negative.
Let $v_{1},\ldots,v_{k}$ be vectors in the same Hilbert space.
We denote the Gram matrix of $v_{1},\ldots,v_{k}$ 
(the matrix made of the inner products
$\langle v_{s}, v_{t} \rangle$) by
$
{\rm Gram}(\{v_{i}\}_{i=1}^{k})$.

Linear operators will be denoted by any of lower case, upper case
or Greek letters. The matrix of transition probabilities of a Markov chain
will usually be denoted by $P$. The typical state space for our Markov 
chains is $[n] = \{1,\ldots,n\}$. Although we avoided 
using $i$ as a running index
in formulas containing the complex root of $-1$, the reader 
needs to use judgment about the meaning of $i$ in each formula.

\section{Bipartite Walks}

\begin{figure}[h]\label{firstfigure}
\centering
\begin{tabular}{ccc}
\epsfig{file=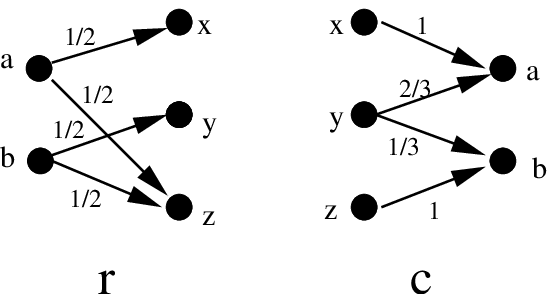} & \hspace{1in} & 
\epsfig{file= 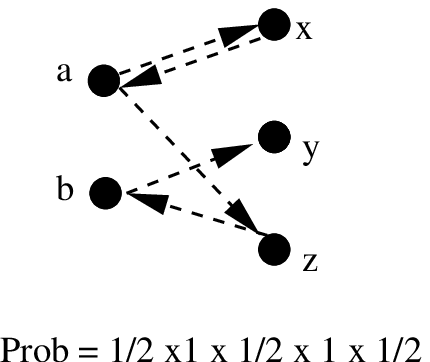} \\
Walk & & Instantiation of length 5\\
\end{tabular}
\caption{Example to a bipartite walk, and one of its instantiations}
\end{figure}

Let $[n|$ and $|m]$ be two disjoint sets of size $n$ and $m$,
respectively. A bipartite walk on $[n|\cup |m]$ is a probabilistic
map with domain $[n|$ and range $|m]$,
and another probabilistic map with domain $|m]$ and range $[n|$.
A probabilistic map is a stochastic matrix with rows and columns
indexed by the elements of the
domain and range of the map, respectively.
Thus a walk is described by a pair of stochastic 
matrices $(\rw,\cl)$ of dimensions $(n,m)$ and $(m,n)$, respectively.
Since $\rw$ and $\cl$ are stochastic, we have that 
$\rw[i,j]\ge 0$ and $r[j,i]\ge 0$
for all
$ 1\le i\le n$, $1\le j\le m$, 
$\sum_{j=1}^{m} \rw[i,j] = 1$ for all $1\le i\le n$, 
$\sum_{i=1}^{n} \cl[j,i] = 1.$ for all $1\le j\le m$. 

To implement the walk we need registers
{\tt Left} and {\tt Right}, that
hold values from $[n|$ and $|m]$, respectively. 
An instantiation of the walk starts at an initial value $a\in [n|$.
We alternately set ${\tt Right}$ to $j$ with probability
$\rw[{\tt Left},j]$ and then
${\tt Left}$ to $i$ with probability $\cl[i,{\tt Right}]$, etc.
If $a$ comes from an initial distribution on $[n|$, then
executing the walk for an even number of steps 
results in
a distribution on $[n|$, and executing the walk for an odd number of steps 
results in
a distribution on $|m]$.

The quantized version of the walk
takes place in the $n\times m$
dimensional Hilbert space of quantum registers
$|{\tt Left}\rangle |{\tt Right}\rangle$.
The roles of $\rw$ and $\cl$ are taken up by diffusion operators
$2C - I $ and $2R - I$. Operator $2C - I$ is controlled by the first register
and acts on the second register. Let $\rw_{i}$,
($1\le i\le n$) be the vector of
probabilities, where $\rw_{i}[j]$ is the probability that
$\rw$ takes the $i^{\rm th}$ element of $[n|$ into the
$j^{\rm th}$ element of $|m]$, and
let $\cl_{j}$, ($1\le j\le m$)
be the vector of probabilities, where $\cl_{j}[i]$ 
is the probability that $\cl$ takes the $j^{\rm th}$ 
element of $|m]$ into the $i^{\rm th}$ element of $[n|$.
When $|{\tt Left}\rangle$ is set to basis vector $|i\rangle$
we define the controlled diffusion
operator on register $|{\tt Right}\rangle$ as
\[
2\sqrt{\rw_{i}}\sqrt{\rw_{i}}^{T} -I_{|m]}.
\]
The controlled diffusion operator of the $|{\tt Left}\rangle$ register,
when $j$ is the value of the $|{\tt Right}\rangle$ register
is similarly defined by 
\[
2\sqrt{\cl_{j}}\sqrt{\cl_{j}}^{T} -I_{[n|}.
\]
Our focus of interest will be the operator, which does a controlled diffusion 
on $|{\tt Right}\rangle$ and then another one on $|{\tt Left}\rangle$.
We redefine the above using the ket formalism. Let
\begin{eqnarray*}
\brw_{i} & = & \sum_{j=1}^{m} 
\rw[i,j]|i\rangle |j\rangle,\;\;\;\;\; (1\le i\le n); \\
\bcl_{j} & = & \sum_{i=1}^{n} 
\cl[j,i]|i\rangle |j\rangle,\;\;\;\;\; (1\le j\le m).
\end{eqnarray*}
Writing all vectors in the basis
$|i\rangle|j\rangle$ ($1\le i\le n$, $1\le j\le m$) we define the projection
operators
\begin{eqnarray}\label{RWdef}
\RW & = & \sum_{i=1}^{n} \sqrt{\brw_{i}}\sqrt{\brw_{i}}^{T}; \\\label{CLdef}
\CL & = & \sum_{j=1}^{m} \sqrt{\bcl_{j}}\sqrt{\bcl_{j}}^{T}.
\end{eqnarray}

\begin{definition}\label{mudef}
The quantized version of the bipartite walk $(\rw,\cl)$ is
the pair $(2\RW-I,2\CL-I)$ of diffusion operators on
$\C^{[n|\times |m]}$. The two-step walk operator
is $\mu = (2\CL - I)(2\RW-I)$.
\end{definition}

In Section \ref{eigensection} we shall compute
the eigenvalues and eigenvectors of $\mu$
in a more general setting. Equipped with these
expressions in Section \ref{deltaepsilonsec}
we analyze the running time of 
our generalization of Grover/Ambainis 
type algorithms.

\section{The Discriminant Matrix}

The spectrum of $\mu$ of the previous Section is
completely determined by matrices $\rw$ and $\cl$.
But how? The following definition takes us one step closer
to answering this question.
 
\begin{definition}[Discriminant Matrix]
We define the {\em discriminant matrix} of a bipartite walk
$(\rw,\cl)$ as:

\begin{equation}\label{Mexpress}
M = \left(\begin{array}{ll} 
0 & \sqrt{\rw\circ \cl^{T}} \\
\sqrt{\cl\circ \rw^{T}} & 0 \\
\end{array}\right).
\end{equation}

\end{definition}

The significance of the discriminant matrix will 
become clear from our Spectral Theorem. Below we give an 
equivalent definition of it, 
that will yield itself to a natural generalization.
Notice that:

\begin{eqnarray}\label{ort1}
\langle \sqrt{\bf \rw_{i}} \;|\; \sqrt{\bf \rw_{i'}}\rangle 
& = &\delta_{i,i'}\;\;\;\;\; \mbox{for all $1\le i\le i'\le n$; } \\
\label{ort2}
\langle \sqrt{\bf \cl_{j}}\; |\; \sqrt{\bf \cl_{j'}}\rangle 
&  = &\delta_{j,j'}\;\;\;\;\; \mbox{for all $1\le j\le j'\le m$.} 
\end{eqnarray}
Furthermore:
\begin{eqnarray*}
\langle \sqrt{\bf \rw_{i}} \;|\; \sqrt{\bf \cl_{j}}\rangle 
= \sqrt{\rw[i,j]}\sqrt{\cl[j,i]} \; = \;
M[i,j]. \\
\end{eqnarray*}
Thus we can also express $M$ as:
\[
M = {\rm Gram}(\sqrt{\bf \rw_{1}},\ldots,\sqrt{\bf \rw_{n}},\;\;
\sqrt{\bf \cl_{1}},\ldots,\sqrt{\bf \cl_{m}}) - I_{n+m}.
\]

\section{A Spectral Theorem}\label{spectralsection}

Let $H$ be a Hilbert space and $\A$ be a subspace of $H$.
The unitary operator that 
that leaves $\A$ invariant and takes
all vectors in $\A^{\perp}$ to their opposite is called a general
reflection, and it is denoted by $\rf_{\A}$.
Let $\A, \B \le H$ be defined via two separate
orthogonal bases of unit vectors:
\begin{eqnarray*}
\A & = & \langle {\bf v_{1}},\ldots,{\bf v_{n}} \rangle; \\
\B & = & \langle {\bf w_{1}},\ldots,{\bf w_{m}} \rangle.
\end{eqnarray*}
In Section \ref{eigensection} we show the easy fact that for
$\RW = \sum_{i=1}^{n}{\bf v_{i}}{\bf v_{i}}^{\ast}$ and
$\CL = \sum_{j=1}^{m}{\bf w_{j}}{\bf w_{j}}^{\ast}$ the operators
$2\RW-I$ and $2\CL-I$ are exactly $\rf_{\A}$ and $\rf_{\B}$.
Expressions (\ref{RWdef}), (\ref{CLdef}) and relations 
(\ref{ort1}) and (\ref{ort2}) tie the problem of computing 
the spectrum and eigenvectors of $\mu$ 
in Definition \ref{mudef} to the problem of
computing the spectrum of operator
$\mu = \rf_{\B}\rf_{\A}$. 

\begin{definition}
We call $\A^{\perp} \cap  \B^{\perp}$ the 
{\em idle} subspace, and its orthogonal complement 
the {\em busy} subspace.
\end{definition}

The justification for the above definition is that
the idle subspace lies in the kernel of operators $\RW$,
$\CL$, and hence $\mu$ acts in the idle subspace as the identity.
Thus in order to get the spectral decomposition of $\mu$
it is enough to compute it restricted on the busy subspace.

\begin{definition}
For $a=(\alpha_{1},\ldots,\alpha_{n}|\in \C^{[n|}$ and
$b=|\beta_{1},\ldots,\beta_{m})\in \C^{|m]}$ 
we define 
\begin{eqnarray*}
\tilde{a} & = &
\sum_{i=1}^{n}\alpha_{i} {\bf v}_{i} \\
\tilde{b} & = &
\sum_{j=1}^{m} \beta_{j} {\bf w}_{j} \\
(a,b)^{\sim} & = & \tilde{a} + \tilde{b} = 
\sum_{i=1}^{n}\alpha_{i} {\bf v}_{i} + 
\sum_{j=1}^{m} \beta_{j} {\bf w}_{j}.
\end{eqnarray*}
\end{definition}

Every vector in the busy subspace has
a convenient (albeit not unique) expression using the tilde:
\[
(\A^{\perp} \cap  \B^{\perp})^{\perp} = \langle \A,\B\rangle
= \A+\B = \{\tilde{a} + \tilde{b}\mid\; a\in \C^{[n|}, b\in \C^{|m]}\}. 
\]

We also need:

\begin{definition}[Discriminant Matrix (generalized)]\label{gendisc}
The {\em Discriminant Matrix} of an ordered pair
$(\{ {\bf v_{1}},\ldots,{\bf v_{n}} \} \; ,
\; \{ {\bf w_{1}},\ldots,{\bf w_{m}} \})$ 
of orthonormal systems is:
\[
M= {\rm Gram}({\bf v_{1}},\ldots,{\bf v_{n}},{\bf w_{1}},\ldots,{\bf
  w_{m}}) -I
\]
\end{definition}

Our Spectral Theorem
relates the spectrum and eigenvectors of
$\mu$ on the busy subspace
to the spectrum and eigenvectors of $M$.
The eigenvalues of $M$ are symmetric to $0$ 
and are in the $[-1,1]$ range.


\begin{theorem}[Spectral Theorem]\label{mainth}
Let ${\bf v_{1}},\ldots,{\bf v_{n}}$ and 
${\bf w_{1}},\ldots,{\bf w_{m}}$
be two orthogonal systems of unit vectors
spanning spaces $\A$ and $\B$, respectively. The eigenvectors 
and eigenvalues of the operator $\mu = \rf_{\B}\rf_{\A}$
on $\A + \B$ are derived from those of the
discriminant matrix $M$ of the pair $(\{ {\bf v}_{i} \}_{i\in [n|},
\{ {\bf w}_{j} \}_{j\in |m]})$ as follows:
\medskip

\begin{tabular}{|c|p{4.2in}|}\hline
Eigenvalue & Eigenvector/Space \\\hline\hline
\begin{tabular}{c}
 \\
1 \\
\\
\end{tabular}
& All vectors 
in
$
\A\cap\B.
$
This space has dimension $d_{1}$, where $d_{1}$ is dimension
of the eigen-space of $M$ associated with eigenvalue 1.
\\\hline
\begin{tabular}{c}
 \\
$2\lambda^{2} - 1 - 2i \lambda  \sqrt{1-\lambda^{2}}$ \\
\end{tabular}
&
All vectors of the form
$
\tilde{a} - \lambda \; \tilde{b} + \, i\, 
\sqrt{1- \lambda^{2}}\; \tilde{b},
$
where $(a,b)$ is an eigenvector of $M$ with eigenvalue $1>\lambda>0$.
\\\hline
 & \\
$2\lambda^{2} - 1 + 2i \lambda  \sqrt{1-\lambda^{2}}$ &
All vectors of the form
$
\tilde{a} - \lambda \; \tilde{b} - \, i\, 
\sqrt{1- \lambda^{2}}\; \tilde{b},
$
where $(a,b)$ is an eigenvector of $M$ with eigenvalue $1>\lambda>0$.
\\\hline
 & \\
$-1$ &
All vectors of the form $\tilde{a}$ or $\tilde{b}$, 
where $(a,b)$ is an eigenvector of $M$ with eigenvalue $0$.
(This implies that $(a,0)$ and $(0,b)$ are eigenvectors of 
$M$ with eigenvalue $0$ too.) \\\hline
\end{tabular}
\end{theorem}


\begin{figure}[t]\label{secfigure}
\centering
\epsfig{file= 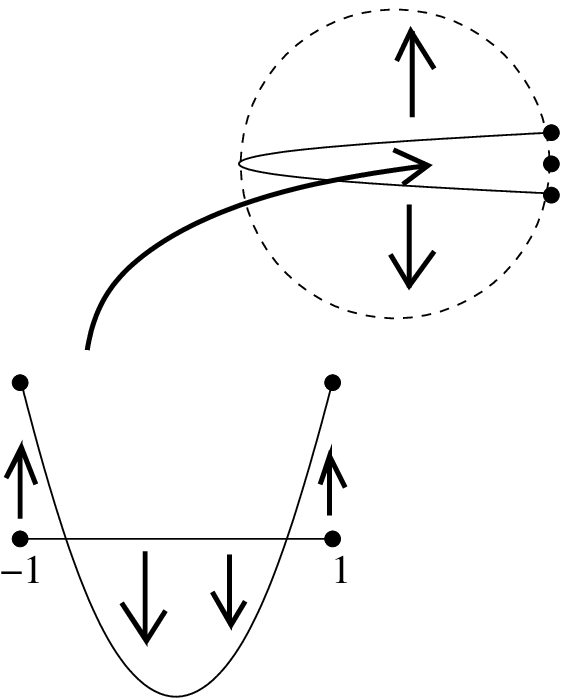}
\caption{The transformation that takes the
eigenvalues of $M$ into the eigenvalues of $\mu$}
folds the $[-1,1]$ interval and then expands it to 
the complex unit circle (dotted circle).
\end{figure}

We prove the Spectral Theorem and other 
unproven claims of this section in Section \ref{eigensection}.
Let $(a,b)\in {\bf C}^{[n|\cup |m]}$.
When $\{{\bf v_{i}}\}_{i\in [n|}$
and $\{{\bf w_{i}}\}_{j\in |m]}$ specialize to $\{\sqrt{\rw_{i}}\}_{i}$
and $\{\sqrt{\cl_{j}}\}_{j}$, the tilde operator
takes the following form:
\begin{eqnarray*}
(a,b)^{\sim} = \sum_{i=1}^{n} \alpha_{i} \; 
|i\rangle \left( \sum_{j=1}^{m} M[i,j]\; |j\rangle \right)
+\sum_{j=1}^{m} \beta_{j} \; \left(\sum_{i=1}^{n} 
 N[j,i]\; |i\rangle \right) \; |j\rangle & = & \\
\sum_{i=1}^{n} \sum_{j=1}^{m} (\alpha_{i} M[i,j] + \beta_{j} N[j,i]) \;
|i\rangle |j\rangle.
\end{eqnarray*}
The busy subspace is this case is the set of all vectors that can be
expressed as above, and the discriminant matrix specializes to 
(\ref{Mexpress}).

\section{Bipartite Walks from Ordinary Markov Chains}
\label{walksfromchains}

A Markov chain with state set $[n]$
is an $n$ by $n$ stochastic 
matrix $P$. We associate the (classical) bipartite walk $(P,P)$ with 
chain $P$. The discriminant matrix of the walk is

\begin{equation}\label{markovmtr}
M = \left(\begin{array}{ll} 
0 & D \\
D & 0 \\
\end{array}\right),
\end{equation}
where $D = \sqrt{P\circ P^{T}}$. We call $D$ the 
{\em half discriminant matrix}. Since $M = 
\left(\begin{array}{ll}
0 & 1 \\
1 & 0 
\end{array}\right)
\otimes D$, its eigenvectors are of the form $(a,a)$ and $(a,-a)$,
where $a$ is an eigenvector of $D$.
Matrix $D$ in general is not stochastic, and
it can be the zero matrix, for instance when $P$ is the matrix
of a cyclic permutation.
A stationary distribution of a Markov chain $P$ 
is a probability distribution $\varpi$ on $[n]$ such that $\varpi P = \varpi$.
It is an eigenvector of $P$ with all non-negative components.
A Markov chain $P$ is {\em symmetric} if $P=P^{T}$. 
In this case $D=P$ and its matrix is doubly stochastic. 
The uniform distribution on $[n]$
is stationary for symmetric Markov chains.
In the quantized case:

\begin{lemma}\label{qstationary}
Let $P$ be a symmetric Markov chain on $[n]$ and
let $\mu$ be the quantized walk operator associated with $P$.  
Let $a = ({1\over\sqrt{n}},\ldots,{1\over\sqrt{n}}|$,
$b = |{1\over\sqrt{n}},\ldots,{1\over\sqrt{n}})$.
Then $u = \tilde{a} = \tilde{b}$ is a unit vector
and $\mu u = u$.
\end{lemma}

\begin{proof}
We have
\begin{eqnarray}
u = {1\over\sqrt{n}}
\sum_{i=1}^{n} \sqrt{\brw_{i}} = \sum_{i=1}^{n}{1\over\sqrt{n}}
\sum_{j=1}^{m}\sqrt{P[i,j]} \; |i\rangle|j\rangle
& = & \\\label{ueq}
\sum_{1\le i,j\le n} \sqrt{P[i,j]\over n}\; |i\rangle|j\rangle
 & = & \\
\sum_{j=1}^{m} {1\over\sqrt{n}}\sum_{i=1}^{n} \sqrt{P[j,i]}\;
|i\rangle|j\rangle
= {1\over\sqrt{n}}
\sum_{j=1}^{m} \sqrt{\bcl_{i}}. & & 
\end{eqnarray}
From Lemma \ref{intersectlemma} it follows that
$u$ is an eigenvector of $\mu$ 
with eigenvalue 1. Direct calculation or
Theorem \ref{normth} gives $|u|=1$.
\end{proof}

\section{Walks with ``Memory''}\label{memorysec}

We first study the classical setting.
Let $P$ be a symmetric Markov chain with state set $[n]$.
Assume furthermore that elements of a subset $\gig $ of $[n]$ are marked, where
$\gig $ is either the empty set or
$|\gig| \ge \epsilon n$.
In this section we consider the task 
of designing an efficient algorithm that 
differentiates in between the above two cases.
At our disposal there are three subroutines,
each with different associated costs:

\bigskip

\begin{tabular}{l|p{3.5in}|c}
Name of the routine & Description & Cost \\\hline
{\tt PickUniform()} &  Picks a random $i\in [n]$ distributed according to $\varpi$ &
$\wp_{0}$ \\
{\tt ApplyChain(i)} &  Applies the randomized map $P$ on input $i\in [n]$ &
$\wp_{1}$ \\
{\tt IsMarked(i)} & Returns $0$ if $i\in [n]\setminus \gig $, returns
$1$ if $i\in \gig $ & $\wp_{2}$ \\
\end{tabular}
\bigskip

\noindent Under the above conditions what is the price of
completing the task? The optimal algorithm depends on 
the ratios of $\wp_{0}$, $\wp_{1}$, $\wp_{2}$. If 
$\wp_{0}$, $\wp_{1}$, $\wp_{2}$ have the same magnitude
we just repeatedly pick random elements of $[n]$ 
using {\tt PickUniform()} and test them
with {\tt IsMarked(i)} until we either find a marked element
or declare that $\gig $ is empty.
We need to do this $O(1/\epsilon)$ times to achieve
small constant error probability.
However, if $\wp_{1}$ and $\wp_{2}$ are
much smaller than $\wp_{0}$, then the following algorithm 
will perform better:

\bigskip

{\bf Algorithm FindMarked($K$) (Classical):} 
{\tt
\begin{tabbing}
123\=123\=123\=\kill
\> Set i = PickUniform(); \\
\> Do $K$ times \{ \\
\>\> If ( IsMarked(i) == 0 ) i = ApplyChain(i); \\
\> \} \\
\> Output IsMarked(i); \\
\end{tabbing}
}

\noindent If $\gig =\emptyset$, the output of the above algorithm is always
$0$. On the other hand, if 
the eigenvalue gap of $P$ is $\delta$, $|\gig| \ge \epsilon n$ and
$K$ is $1000/\delta\epsilon$, it is easy to see that
{\tt i} will almost certainly "converge" to an element of $\gig $
making the output $1$ with high probability. The associated cost is:
\begin{equation}\label{costeq}
\mbox{Classical cost of detecting large $\gig $} \;\; = \;\; \wp_{0}
+  1000(\wp_{1}+\wp_{2})/(\delta\epsilon).
\end{equation}

We shall "quantize" Algorithm {\tt FindMarked($K$)}
and show that in the quantum case the right hand term of
Expression (\ref{costeq})
is replaced with $1000(\wp_{1}+\wp_{2})/\sqrt{\delta\epsilon}$.
The merit of the result is that both $\epsilon$ and $\delta$ get
square rooted, while what (almost) trivially follows from Grover
is only that $\epsilon$ gets square rooted.
Our quantum machine will have registers:

\begin{eqnarray*}
\begin{array}{llll}
\mbox{\bf Control register:} & &
|b\rangle, & \mbox{where $b\in \{0,1\}$;} \\
\mbox{\bf Walk registers:} & &
|i\rangle|j\rangle, &  \mbox{where $i,j\in [n]$.} 
\end{array}
\end{eqnarray*}

We explain the role of the control register later.
The $|i\rangle|j\rangle$ register pair is used to perform 
a quantized bipartite walk with $[n|=|m]= [n]$.
Our new twist is that instead of quantizing $P$, we quantize:
\[
P'\;\stackrel{def}{=}\;\;
\left\{ 
\begin{array}{l}
\mbox{if $i\not\in \gig $ use map $P$ on $i$;} \\
\mbox{if $i\in \gig $ then map $i$ into itself with probability 1.}
\end{array}\right.
\]
We might say that $P'$ ``remembers'' if it ever sees a marked element.
If $\gig =\emptyset$
then $P'=P$. For the general case, express:
\[
P = 
\left(
\begin{array}{cc}
P_{1} & P_{2} \\
P_{2}^{T} & P_{3}
\end{array}
\right)
\;\;\;\;
\mbox{with coordinate-division}
\;\;\;\;\;\;
\begin{array}{c|cc}
 & [n]\setminus \gig  & \gig  \\\hline
 [n]\setminus \gig  & P_{1} & P_{2} \\
\gig  & P_{2}^{T} & P_{3}
\end{array}
\]
Then the matrix form of $P'$ is:
\[
P' = 
\left(
\begin{array}{cc}
P_{1} & P_{2}\\
0 & I
\end{array}
\right).
\]
Above $I$ is the identity matrix. 
The half discriminant matrix 
of the associated bipartite walk is: 
\[
D = 
\left(
\begin{array}{cc}
P_{1} & 0\\
0 & I
\end{array}
\right).
\]
Let us denote by $\nu$ the quantum walk associated with $P'$
and by $\mu$ the quantum walk associated with $P$. Define
\[
u = \sum_{1\le i,j\le n} \sqrt{P[i,j]\over n}\; |i\rangle |j\rangle.
\]
By Lemma \ref{qstationary} and (\ref{ueq})
we have that $\mu u = u$. Therefore
$\mu^{K} u = u$ for every integer $K$. 
Our algorithm will utilize:

\begin{enumerate}
\item If $\gig =\emptyset$ then $\nu=\mu$, so $\left|{u + \nu^{K}u
\over 2}\right| = |(u+u)/2| = 1$.
\item If $|\gig| \ge \epsilon n$ 
and $K\ge 1000/\sqrt{\delta\epsilon}$ then
$
\left|{u +  \nu^{K} u\over 2}\right| \le 3/4.
$
\end{enumerate}

\noindent The above formula explains the role of the control register.
At the start we split the computation into two branches:
in one branch we leave $u$ untouched while in the
other we apply $\nu^{K}$ on it. The splitting operator is
$H = (= H^{-1}) = {1\over \sqrt{2}} \left(\begin{array}{rr} 
1 & 1 \\ 1 & - 1 \end{array} \right)$.
The algorithm: 
\bigskip

{\bf Algorithm FindMarked($K$) (Quantum):} 
{\tt
\begin{tabbing}
123\=123\=123\=\kill
\> Put the pair of walk registers into the state $u$; \\
\> Apply $H$ on the Control register; \\
\> /$\ast$ This puts the system into state ${1\over\sqrt{2}}|0\rangle u + 
{1\over\sqrt{2}}|1\rangle u$; $\ast$/ \\
\> Do $K$ times \{ \\
\>\> If ( the Control register is $|1\rangle$ ) \{ \\
\>\> apply $\nu$ 
on the pair of walk registers;  \\
\> \} \\
\> /* Now the machine will be at state 
${1\over\sqrt{2}}|0\rangle u + 
{1\over\sqrt{2}}|1\rangle \nu^{K} u$ */ \\
\> Apply $H^{-1}$ on the Control register; \\
\> Measure the final state: $|b\rangle|i\rangle|j\rangle$; \\
\> If ( b = 1 or $i\in \gig $ ) output 1, else output 0.
\end{tabbing}
} 

Next we give a pair of subroutines that
put the walk register $|i\rangle|j\rangle$ into the
state $\nu |i\rangle|j\rangle $. These routines
implement the operators $2\RW'-I$ and $2\CL'-I$, where
$(2\RW'-I,2\CL'-I)$ is the quantization of Markov chain $P'$.
At our disposal, like in the classical analogue, 
we only have operators $2\RW-I$ and $2\CL - I$,
where $(2\RW-I,2\CL-I)$ is the quantization of Markov chain $P$.
While $P$ does not not depend on the marked subset,
$P'$ does. Thus our subroutines will need to use subroutine {IsMarked(i)}.
They will also use an ancilla register $|a\rangle$
named Marked, where $a\in\{0,1\}$. This register stays 
$|0\rangle$ in between applications of the subroutines
and is not to be confused with the Control register.
\bigskip

{\bf Algorithm PerturbedWalk, diffusion on $j$:} 
{\tt
\begin{tabbing}
123\=123\=123\=\kill
\> $|i\rangle|j\rangle|0\rangle \rightarrow
|i\rangle|j\rangle|g(i)\rangle$, where g(i) = IsMarked(i); \\
\> If $g(i)=0$ apply $(2\RW-I)$ on the pair of walk registers; \\
\> /* This takes $|i\rangle|j\rangle|g(i)\rangle$ for g(i)=0
into the state \\
\>\> $|i\rangle \left( \sum_{j'\in |n]} 2\sqrt{P'[i,j']} |j'\rangle
\right) |g(i)\rangle
- |i\rangle |j\rangle |g(i)\rangle  $ */ \\
\>   $|i\rangle|j\rangle|a\rangle \rightarrow
|i\rangle|j\rangle|a\oplus g(i)\rangle$, where g(i) = IsMarked(i); \\
\> /* This step ``forgets'' g(i), since a=g(i) at this stage */ \\
\end{tabbing}
}

\noindent Similarly:

\medskip
 
{\bf Algorithm PerturbedWalk, diffusion on $i$:} 
{\tt
\begin{tabbing}
123\=123\=123\=\kill
\> $|i\rangle|j\rangle|0\rangle \rightarrow
|i\rangle|j\rangle|g(j)\rangle$, where g(j) = IsMarked(j); \\
\> If $g(j)=0$ apply $(2\CL-I)$ on the pair of walk registers; \\
\> /* This takes $|i\rangle|j\rangle|g(j)\rangle$
for g(j)=0 into the state \\
\>\> $\left( \sum_{i'\in [n|} 2 \sqrt{P'[j,i']} |i'\rangle
\right) |j\rangle |g(j)\rangle
-|i\rangle |j\rangle |g(j)\rangle$ */ \\
\>   $|i\rangle|j\rangle|a\rangle \rightarrow
|i\rangle|j\rangle|a\oplus g(j)\rangle$, where g(j) = IsMarked(j); \\
\> /* This step ``forgets'' g(j), since a=g(j) at this stage */ \\
\end{tabbing}
}

In the quantized setting we need to reinterpret costs $\wp_{0}$
and $\wp_{1}$: The cost of putting the machine into state $u$
is $\wp_{0}$; The cost of applying operators 
$2\RW-I$ and $2\CL-I$ on the pair of walk registers 
is $\wp_{1}$.
With this pricing the total cost of the operation FindMarked($K$) is 
\[
O(\wp_{0} + K(\wp_{1} + \wp_{2})).
\]

\section{The $\sqrt{\delta\epsilon}$ rule}\label{deltaepsilonsec}

\begin{theorem}\label{deltaepsilonth}
Let $P$ be a symmetric Markov chain with state set $[n]$,
and let $\gig $ be a subset of $[n]$ marked via
operator {\tt IsMarked(i)}. Let $\delta$ be the eigenvalue gap 
of $P$. Then for a randomly picked
$K\in [1,1000/\sqrt{\delta\epsilon}]$:
\[
\begin{array}{lll}
1. & \mbox{If $\gig =\emptyset$} & \mbox{FindMarked($K$)
outputs $0$ with probability $1$;} \\
2. & \mbox{If $|\gig |\ge \epsilon n$} & \mbox{FindMarked($K$)
outputs $1$ with probability 
at least $1/1000$}.
\end{array}
\]
(The probabilities are both over $K$ and the output of 
FindMarked($K$).)
\end{theorem}

\begin{proof}
Recall that state of the machine during the execution
of FindMarked($K$) evolves as:
\begin{eqnarray}\label{lasterr}
|0\rangle 0 \;\rightarrow\;
|0\rangle u \;\rightarrow\;
{1\over \sqrt{2}}|0\rangle u + {1\over\sqrt{2}}|1\rangle u \; 
& \rightarrow & \\
{1\over \sqrt{2}}|0\rangle u + 
{1\over\sqrt{2}}|1\rangle \nu^{K} u\; & \rightarrow & \\
|0\rangle {u + \nu^{K} u\over 2} + |1\rangle {u - \nu^{K} u\over 2}.
\end{eqnarray}
The proof of the last transition:
\begin{eqnarray}\nonumber
\left( {1\over\sqrt{2}} | 0\rangle u +{1\over\sqrt{2}} | 1\rangle
\nu^{K} u\right)
\left(
\begin{array}{cc}
{1\over\sqrt{2}} & {1\over\sqrt{2}} \\\label{twelve}
{1\over\sqrt{2}} & - {1\over\sqrt{2}}
\end{array}\right) \otimes I_{{\bf C}^{[n|\times n]}} & = & \\
{1\over 2}| 0 \rangle (u+ \nu^{K} u) + {1\over 2}| 1\rangle
(u- \nu^{K} u).
\end{eqnarray} 
The last arrow of (\ref{lasterr})
represents a transition (actually the only one)
that depends on the input, since $\nu$, the quantized 
version of the modified walk, $P'$, depends on what $\gig $ is.
If $\gig =\emptyset$ then $\nu = \mu$ (see previous section),
and since $u$ is left invariant under $\mu$, 
the final state is $|0\rangle u$, and the output 
is 0 with probability 1.
This proves the first part of
theorem. 

In order to prove the second part
we show that if $|\gig| = \epsilon n$, then
\begin{equation}\label{eqtoshow}
\left|{u + \nu^{K} u\over 2}\right|\le {7\over 8}
\end{equation}
or the left walk register contains an element of 
$\gig$ with probability $1/16$. In fact, if 
$\epsilon > 1/8$ the latter is the case.
To see this notice that the last operation does not 
effect the walk register, thus we get the same measurement
for the walk register as if we measured it in the previous step.
Taking a look now at Formula $(\ref{twelve})$, and the fact that
$u$ us ``uniform,'' we can easily see the claim. 
Therefore in the sequel we shall
assume that $\epsilon \le 1/8$.

We use Theorem \ref{mainth} to compute the spectral decomposition of
$\nu$ and in turn to compute the effect of $\nu^{K}$ on $u$.
Recall that $\nu$ is the quantization of the chain $P'$,
so in order to apply Theorem \ref{mainth} we first need to 
determine the eigenvalues and eigenvectors of 
the half discriminant matrix
\[
D = \sqrt{P'\circ {P'}^{T}} = \left(
\begin{array}{cc}
P_{1} & 0\\
0 & I
\end{array}
\right).
\]
For $i\in [n]$ let $e_{i}$ be the unit vector that 
takes $1$ on $i$ and $0$ elsewhere.
Because of the block structure of $D$, for all $i\in \gig $ the vector
$e_{i}$ is an eigenvector of $D$. The remaining 
eigenvectors are those
that are eigenvectors of $P_{1}$ augmented with zeros
on the coordinates corresponding to the elements of $\gig $.

\begin{lemma}\label{radiuslemma}
The spectral radius of $P_{1}$ is at most $1-\delta\epsilon/2$.
\end{lemma}

\begin{proof}
It is an easy fact that
there exists a non-negative unit vector $\rho'$ such that
$\rho' P_{1} = \lambda'\rho' $ and $\lambda'$ is the 
spectral radius of $P_{1}$. Define $\rho$ to be the
augmentation of $\rho'$ with zeros
on the coordinates corresponding to the elements of $\gig $,
and define $\vc = \sum_{i=1}^{n} {1\over \sqrt{n}} e_{i}$.
Since $P$ is doubly stochastic we have $\vc P = \vc$. Consider
the spectral decomposition of $\rho$
in the basis formed 
by the eigenvectors of $P$:
\[
\rho  = \alpha \vc  + \sum_{k} \alpha_{k} \vc_{k},
\]
where $\{\vc_{k}\}\cup\{\vc \}$ is a complete set of eigenvectors
for $P$. Since $P$ is symmetric, its eigenvectors are orthogonal
and we have $\alpha^{2} + \sum_{k}\alpha_{k}^{2} = 1$.
For a vector
$v\in {\bf C}^{[n]}$ we denote by $v_{G}$ and $v_{\overline{G}}$
its $G$ and $[n]\setminus G$ components. $|G|\ge \epsilon n$
is equivalent to:
\[
\langle a_{\overline{G}} , a_{\overline{G}} 
\rangle = 
\langle a_{\overline{G}} , a \rangle \le 1-\epsilon.
\]
Since $\rho_{G} = 0$:
\begin{eqnarray*}
\alpha = 
\langle \rho , \vc \rangle 
= \langle \rho , \vc_{G} + \vc_{\overline{G}} \rangle =
\langle \rho , \vc_{\overline{G}}\rangle \le 
|\rho ||\vc_{\overline{G}}| \le \sqrt{1-\epsilon}.
\end{eqnarray*}
Hence, $\sum_{k} \alpha_{k}^{2} = 1-\alpha^{2} \ge \epsilon$.
Let $\lambda_{k}$ be the eigenvalue of $P$ associated with $\vc_{k}$.
Then $\rho P = \alpha \vc  + \sum_{k} \alpha_{k}\lambda_{k} \vc_{k}$.
Since the eigenvalue gap of $P$ is at least $\delta$, we have
\begin{eqnarray*}
|\rho P|^{2} = \alpha^{2} + \sum_{k} \alpha_{k}^{2}\lambda_{k}^{2}
& \le & \\
\alpha^{2} + (1-\delta)^{2}\sum_{k} \alpha_{k}^{2} 
& \le & \\
\alpha^{2} + (1-\delta)\sum_{k} \alpha_{k}^{2} 
& = & \\
\alpha^{2} + \sum_{k} \alpha_{k}^{2} - 
\delta\sum_{k} \alpha_{k}^{2} & \le & 1 - \delta\epsilon.
\end{eqnarray*}
On the other hand 
\[
|\rho P|^{2}\ge |\rho D|^{2} = |\rho' P_{1}|^{2} = {\lambda'}^{2}
\]
which implies
${\lambda'}^{2}\le 1- \delta\epsilon$, $\lambda'\le 1-\delta\epsilon/2$,
as needed.
\end{proof}

From the above lemma one can anticipate the proof of the theorem:
Any eigenvector of $\nu$
that comes from the "quantization" of a shrinking eigenvector of $P'$
has eigenvalue $e^{i\theta}$, where $\theta$
is separated away from 0 by at least a constant factor times
$\sqrt{\delta\epsilon}$. This we deduce from the previous lemma and 
Theorem \ref{mainth}.
Hence, if $K$ is a random number in the range 
$[1,1000/\sqrt{\delta\epsilon}]$,
the phase shift, when applying $\nu^{K}$ on any of the 
above eigenvectors
is some non-zero constant times $\pi$ on expectation.
Among the technical details we need to
work out the most compelling (although not very hard)
is that the projection of $u$
on the space spanned by the above eigenvectors is large.
Define
\[
u' = \sum_{i\in [n]\setminus \gig;\; j\in [n]} \;
\sqrt{P[i,j]\over n}\; |i\rangle |j\rangle.
\]
From our assumption that $|G|/n \le {1\over 8}$ we obtain:
\begin{eqnarray*}\label{componenteq}
\langle u,u'\rangle = \; \sum_{i\in [n]\setminus \gig;\; j\in [n]}
{P[i,j]\over n}\;  =\; {n-|G|\over n} \ge {7\over 8}.
\end{eqnarray*}
Since $\nu$ is unitary so is $\nu^{K}$, therefore
\[
| \nu^{K} u -  \nu^{K} u'|^{2} = |\nu^{K}(u-u')|^{2} = |u-u'|^{2} =
|u|^{2}+|u'|^{2} - \langle u,u'\rangle - \langle u',u\rangle
\le  {1\over 4}.
\]
From this
\[
\left|{u+ \nu^{K} u \over 2}\right|^{2} \le
{1\over 4} |u'+ \nu^{K} u'|^{2} + {1\over 4} |u-u'|^{2} +
{1\over 4} |\mu^{K}(u-u')|^{2} 
\le {1\over 4} |u'+ \nu^{K} u'|^{2} + {1\over 8}.
\]
Set
\[
{\rm amp}_{K} \stackrel{def}{=} |u'+ \nu^{K} u'|^{2}.
\]
We are done if we show that 

\begin{lemma}
The probability that
for a random $K\in [1,1000/\sqrt{\delta\epsilon}]$
the value of ${\rm amp}_{K}$ is at most 3 is at least $1/6$.
\end{lemma}

\begin{proof}
We need to set up the stage to use Theorem \ref{mainth}
for the operator $\nu$. Since $\nu$ is the quantization of $P'$, the
busy subspace will be $\A' + \B'$, where $\A'$ is generated by
\begin{eqnarray*}
{\bf v_{i}} = 
\sum_{j\in |n]}  \sqrt{P[i,j]}\; & |i\rangle |j\rangle & \;\;\;\;\;
\mbox{for $i\in [n]\setminus \gig $;}  \\
& |i\rangle |i\rangle & \;\;\;\;\; \mbox{for $i\in \gig $.}
\end{eqnarray*}
and $\B'$ is generated by
\begin{eqnarray*}
 {\bf w_{j}} = \sum_{i\in [n|} \sqrt{P[j,i]}\; & |i\rangle |j\rangle &
\;\;\;\;\;
\mbox{for $j\in [n]\setminus \gig $;} \\
& |i\rangle |i\rangle & \;\;\;\;\; \mbox{for $i\in \gig $.}
\end{eqnarray*}
The space spanned by the vectors 
$|i\rangle |i\rangle$ ($i\in \gig $) is invariant under $\nu$
(from Lemma \ref{intersectlemma}, for instance)
and it is orthogonal to 
$u'= \sum_{i\in [n]\setminus \gig } {1\over\sqrt{n}} {\bf v_{i}}$.
Define $\A'' = \langle {\bf v_{i}} \mid \; 1\le i\le n \rangle$
and $\B'' = \langle {\bf w_{j}} \mid \; 1\le j\le n \rangle$.
Let $Z$ be the orthogonal complement of the subspace
generated by $|i\rangle |i\rangle$ ($i\in \gig $). We have:
\begin{eqnarray}\label{uprime1}
u' & \in & Z \\\label{uprime2}
\nu|_{Z} & = & (\rf_{\B''} \rf_{\A''})|_{Z}.
\end{eqnarray}
The discriminant matrix associated with the 
$(\{{\bf v_{i}}\}_{i\in [n]\setminus G},
\{ {\bf w_{j}}\}_{j\in [n]\setminus G})$
pair is
\[
M_{1} = \left(
\begin{array}{cc}
0 & P_{1} \\
P_{1} & 0
\end{array}
\right).
\]
It follows from Lemma \ref{radiuslemma} that all eigenvalues 
of $M_{1}$ are less than $1-\delta\epsilon/2$ in absolute value
(the eigenvalues of $M_{1}$
and those of $P_{1}$ coincide up to a sign).
Since the action of $\nu$ on $Z$ is the
same as that of $\rf_{\B''} \rf_{\A''}$ and $u'\in Z$, 
we decompose $u'$
according to orthogonal unit eigenvectors of $\rf_{\B''}\rf_{\A''}$:
\[
u' = \sum_{k=1}^{\ell} \gamma_{k} \vvv_{k}.
\]
Since $u'\in \A''$, it
lies in the busy subspace of operators $\rf_{\A''}$ and $\rf_{\B''}$.
We use Theorem \ref{mainth} to claim that each $\vvv_{k}$
on the busy subspace has eigenvalue 
\[
e^{i\theta_{k}} \;\; \stackrel{def}{=} \;\;
2\lb_{k}^{2}-1 + 2i\lb_{k} \sqrt{1-\lb_{k}^{2}},
\]
where $\lb_{k}$ is some eigenvalue of $M_{1}$.
Consequentially
$|\lb_{k}|\le 1-\delta\epsilon/2$. Hence
\[
|\cos\theta_{k}| \;  = \; 2\lb_{k}^{2}-1 \; < \;
2(1- \delta\epsilon/2) -1 \; = \;
1 - \delta\epsilon.
\]
Therefore when representing angles in $[-\pi,\pi]$:
\begin{equation}\label{degreerestrict}
|\theta_{k}| \ge \sqrt{\delta\epsilon}\;\;\;\; \mbox{for $1\le k\le \ell$.}
\end{equation}
We have 
\begin{eqnarray}\nonumber
\nu^{K}u' & = & \sum_{k=1}^{\ell} 
\gamma_{k} e^{i\theta_{k} K} \vvv_{k}
\\\nonumber
u' +  \nu^{K}u' & = & \sum_{k=1}^{\ell}
\gamma_{k} (1 + e^{i\theta_{k} K}) \vvv_{k} \\\label{eq20}
\amp_{K} = |u' +  \nu^{K} u'|^{2} & = & \sum_{k=1}^{\ell}
|\gamma_{k}|^{2} |1 + e^{i\theta_{k} K}|^{2}.  
\end{eqnarray}

In Equation \ref{eq20} we used the fact that unitary 
operators have orthogonal eigenvector-systems.
We show that 
for an individual $k$ the expected value  of $|1 + e^{i\theta_{k} K}|^{2}$
for a random $K\in [0,1000/\sqrt{\delta\epsilon}]$ is close to 2.
Let us denote $1000/\sqrt{\delta\epsilon}$ by $N$. Then:
\begin{eqnarray*}
{1\over N}\sum_{K=1}^{N} |1 + e^{i\theta_{k} K}|^{2} &=& \\
{1\over N}\sum_{K=1}^{N} 2 + e^{i\theta_{k} K} + e^{-i\theta_{k} K}
&=& \\
2 + {1\over N}\sum_{K=1}^{N}  e^{i\theta_{k} K} + {1\over N}\sum_{K=1}^{N}
e^{- i\theta_{k} K} & = & \\
2  + {e^{i\theta_{k} (N+1)} - 1 \over N (e^{i\theta_{k}} - 1)}
+ {e^{- i\theta_{k} (N+1)} - 1 \over N (e^{- i\theta_{k}} - 1)}
& \le & \\
2 + {2\over  N (e^{i\theta_{k}} - 1)} +  {2\over N( e^{-
i\theta_{k}} -1 )} & \le & 2.5,
\end{eqnarray*}
when $N$ is $1000 /\sqrt{\delta\epsilon}$.
The last inequality comes from Inequality (\ref{degreerestrict})
Let $1-p$ be the probability 
that $\amp_{K} > 3$ for a random $K$. 
Then its expectation is lower bounded by $3-3p$.
Since $2.5\sum_{k=1}^{\ell} \gamma_{i}^{2} = 2.5$ 
is an upper bound on the expectation, we have 
$p\ge 1/6$ as needed.
\end{proof}
\end{proof}

\section{Consequences}

As a first consequence we reprove the result of Ambainis
\cite{amb03} with  a stronger implication:

\begin{theorem}
Let $X$, $Y$ be finite sets,
$f: X\rightarrow Y$ be an oracle function and let 
${\cal R}\subseteq Y\times Y$ be a binary relation known to us.
For $H\subseteq X$ we define $f(H)= \{f(i)\mid \;i\in H\}$.
Define 
\[
p(f,\alpha) = \mbox{ The probability that 
${\cal R}\cap (f(H)\times f(H)) \neq \emptyset$  
for a random set $H$ with size $|X|^{\alpha}$.}
\]
Then there is a quantum query machine 
with oracle $f$ that can differentiate in between the cases
when $p(f,\alpha)=0$ and $p(f,\alpha)\ge \epsilon$ that 
runs in time $O(|X|^{\alpha} + 1000\sqrt{|X|^{\alpha}/\epsilon})$.
\end{theorem}

\begin{proof}
Let $k= \lceil |X|^{\alpha}\rceil$, $n={|X| \choose k}$, and $P$
be the Markov chain on all $k$ subsets of $X$ with transitions:
the probability that $H$ goes to $H'$ is zero
if the the symmetric difference of $H$ and $H'$ is not two,
and ${1\over k(|X|-k)}$ otherwise. We quantize this chain,
but with the caveat that with $H$ we keep track of $f(H)$.
A set $H$ is marked if ${\cal R}\cap (H\times H) \neq\emptyset$.
Since at all times we update $f(H)$, it does not cost us queries
to find out if $H$ is marked. Hence $\wp_{2} = 0$.
The cost to perform $2\RW-I$ or $2\CL- I$ is 2.
Hence $\wp_{1} = 2$.
To put the walk registers into the position
\[
u \;\;= {1\over \sqrt {nk(|X|-k)}} \sum_{|H\Delta H'|=2} |H,f(H)\rangle\;
|H', f(H')\rangle
\]
costs $O(|X|^{\alpha})$ queries. Hence $\wp_{0}= |X|^{\alpha}$.
The probability that an item is marked is 
$p(f,\alpha)$, which is either zero or at least $\epsilon$.
Finally, the eigenvalues of the above Markov chain (derived from
the Johnson graph)
are well known to be ${(k-j)(|X|-k-j)-j\over k(|X|-k)}$ for $j=0,1,\ldots,n$.
Thus its eigenvalue gap is 
\[
\delta = 1-{(k-1)(|X|-k-1)-1\over k(|X|-k)}  = {n\over k(|X|-k)} > 1/k.
\]
The theorem now follows from Theorem \ref{deltaepsilonth}. 
\end{proof}

\begin{corollary}
There is a quantum query machine
running in time $O(|X|^{2/3})$ that differentiates in between the cases
when $f(i)\neq f(i')$ for all $i\neq i'\in X$ and when there are
$i\neq i'\in X$ such that $f(i)= f(i')$.
\end{corollary}

\begin{proof}
Let ${\cal R}$ be the equality relation and set $\alpha=|X|^{2/3}$. 
\end{proof}

\noindent The second consequence of Theorem \ref{deltaepsilonth} is immediate:

\begin{theorem}
Let $G$ be $d$-regular expander on $\{1,\ldots,n\}$. Then
there is a version of Grover search that runs in $\sqrt{n}$ steps
and all transitions are done along the edges of $G$.
\end{theorem}

A special case, where we perform
Grover search along the edges of the hypercube was 
studied by Julia Kempe \cite{k}.

\section{Acknowledgments}

The author thanks M. Santha and F. Magniez for helpful discussions.
Frederick Magniez has introduced me to the result of Ambainis 
at an early stage, and he also showed its relation
with Grover's algorithm.

\newcommand{\etalchar}[1]{$^{#1}$}

\newpage

\section{Operator Notation in Linear Algebra}

This part of the paper builds on ordinary linear algebra.
Correspondingly, every vector (unless transposed)
is a row vector. Operators (as opposed to
the first part) are acting on the right,
and in general we have to reverse the direction of
all formulas of the first part involving vectors, stars and
operators. For inner product we keep the angular notation,
but we separate with comas rather than bars. In particular,
if $v$ is a vector and $M$ and $N$ are linear operators then 
$vMN$ is a row vector that we obtain by applying 
$M$ and $N$ on $v$ in this order. Also:
\begin{eqnarray*}
\langle v,v\rangle =  vv^{\ast} & & \;\;\;\;\;\; \mbox{is a scalar}; \\
v^{\ast} v & &  \;\;\;\;\;\; \mbox{is a 
$\dim v$ by $\dim v$ matrix}.
\end{eqnarray*}

In order to represent elements of ${\bf C}^{[n|}\otimes {\bf C}^{|m]}$ graphically we introduce
{\em Rectangular vectors}. These are vectors with index set
$[n|\times|m]$, drawn in an array format and
delimited by double bars
in order to differentiate them from operators (see figure above). 
Regardless of their 
rectangular shape, they are row vectors in the sense that
operators act on them on the right. If $v$ and $w$ are two rectangular
vectors of the same dimensions, we can take their scalar product
or we can create an operator by writing $v^{\ast}w$ that acts on
${\bf C}^{[n|}\otimes {\bf C}^{|m]}$. A nice thing about 
rectangular vectors is that we can conveniently express the tilde
operator with them.
Let $M$ be an $n$ by $m$ matrix and $N$ be an
$m$ by $n$ matrix. We obtain the rectangular vector ${\bf M}_{i}$ by
replacing all entries of $M$ with zero, except those ones in the
$i^{\rm th}$ row. Similarly, we obtain ${\bf N}_{j}$ by
replacing all entries of $N^{T}$ with zero, except those ones in the
$j^{\rm th}$ column. When $N$ and $M$ are given and fixed, and 
$a = (\alpha_{1},\ldots,\alpha_{n}|$, 
$b = |\beta_{1},\ldots,\beta_{m})$, then
\begin{eqnarray}
\tilde{a} & = & \sum_{i=1}^{n} \alpha_{i} {\bf M}_{i}; \\
\tilde{b} & = & \sum_{j=1}^{m} \beta_{j} {\bf N}_{j}; \\\label{gentilde}
(a,b)^{\sim}  & = & \tilde{a} + \tilde{b}.
\end{eqnarray}
We use the rectangular vector notation only in Section 
\ref{illustration}, where where we work out a
specific example. 

\begin{figure}[t]

\[
\begin{array}{||ccc||c||ccc||}
0 & \sqrt{1\over 3} & 0 & \hspace{0.6in} & 0 & 0 & 0 \\ 
0 & \sqrt{1\over 3} & 0 &  & {1\over 2} & {1\over 2} & {1\over
\sqrt{2}} \\
0 & \sqrt{1\over 3} & 0 &  & 0 & 0 & 0 \\
\end{array}
\]

\caption{Two rectangular unit vectors with inner product
$1/(2\sqrt{3})$.}
\end{figure}

\section{Spectra of Products of two General Reflections}
\label{eigensection}

In this section we prove the claims of Section \ref{spectralsection}.
We need the definitions of that section with the modification
that in this linear algebra inspired part we write operators 
on the right.
Recall that subspaces $\A, \B \le H$ are defined via two separate
orthogonal bases of unit vectors:
\begin{eqnarray*}
\A & = & \langle {\bf v_{1}},\ldots,{\bf v_{n}} \rangle, \\
\B & = & \langle {\bf w_{1}},\ldots,{\bf w_{m}} \rangle,
\end{eqnarray*}
and
\begin{eqnarray*}
\RW & = & \sum_{i=1}^{n} {\bf v}_{i}^{\ast} {\bf v}_{i}, \\
\CL & = & \sum_{j=1}^{m} {\bf w}_{j}^{\ast} {\bf w}_{i}. \\
\end{eqnarray*}

\begin{lemma}\label{imagelemma}
$\RW$ is an orthogonal projection to $\A$ and 
$\CL$ is an orthogonal projection to $\B$.
Also:
\begin{eqnarray}\label{unit1}
\rf_{\A} & = & 2\RW - I; \\
\rf_{\B} & = & 2\CL - I . 
\end{eqnarray}
\end{lemma}

\begin{proof}
We prove that $\RW$ is an orthogonal projection to $\A$ and 
(\ref{unit1}). The other claims are analogous.
For ${\bf v}_{l}$ ($1\le l\le n)$ we have
\begin{equation}\label{ina}
{\bf v}_{l} \RW = \sum_{i=1}^{n} {\bf v}_{l}
{\bf v}_{i}^{\ast} {\bf v}_{i} = \sum_{i=1}^{n} \langle {\bf v}_{l},
{\bf v}_{i}\rangle {\bf v}_{i} = {\bf v}_{l}.
\end{equation}
For any $u$ which is orthogonal to all ${\bf v}_{i}$s
we have:
\begin{equation}\label{orta}
u \RW = \sum_{i=1}^{n} u
{\bf v}_{i}^{\ast} {\bf v}_{i} = \sum_{i=1}^{n} \langle u,
{\bf v}_{i} \rangle {\bf v}_{i} = 0.
\end{equation}
Equations (\ref{ina}) and (\ref{orta})
prove that $\RW$ is an orthogonal projection to $\A$.
Also from (\ref{ina}) and (\ref{orta}):
${\bf v}_{l} (2\RW - I) = 2 {\bf v}_{l} - {\bf v}_{l} = {\bf v}_{l}$
and $u (2\RW - I) = -u$, and (\ref{unit1}) follows.
\end{proof}

Before computing the eigenvalues/vectors of $\mu$
we study the discriminant matrix of Definition \ref{gendisc}.
$M$ has a blocked structure corresponding to the subdivision of
its rows and columns to ${\bf v}$s and ${\bf w}$s.
The two diagonal blocks are 0 and the two off-diagonal blocks are
transposed conjugates of each other.

\begin{lemma}\label{spectrumlemma}
The spectral norm of $M$ is at most $1$. Furthermore, if 
$(a,b)= (\alpha_{1},\ldots,\alpha_{n},\beta_{1},\ldots,\beta_{m})$
is an eigenvector of $M$
with eigenvalue $1$ then 
\[
\sum_{i=1}^{n} \alpha_{i} {\bf v_{i}} = 
\sum_{j=1}^{m} \beta_{j} {\bf w_{j}}.
\]
\end{lemma}

\begin{proof}
Since $M$ is hermitian, all its eigenvalues 
are real.
Let $(a,b) = (\alpha_{1},\ldots,\alpha_{n},\beta_{1},\ldots,\beta_{m})$
be any unit. Then
\begin{eqnarray}\label{cheq}
(a,b)M(a,b)^{\ast} = 2 \sum_{i=1}^{n}\sum_{j=1}^{m} \alpha_{i} 
\overline{\beta_{j}}
\langle {\bf v_{i}},\; {\bf w_{j}} \rangle
= 2 \left\langle\sum_{i=1}^{n} \alpha_{i}{\bf v_{i}},\;\;
\sum_{j=1}^{m} \beta_{j} {\bf w_{j}} \right\rangle.
\end{eqnarray}
Let $\sum_{i=1}^{n} |\alpha_{i}|^{2} = q$, 
$\sum_{j=1}^{m} |\beta_{j}|^{2} = 1-q$. Since
$|\sum_{i=1}^{n} \alpha_{i}{\bf v_{i}}|=\sqrt{q}$ and
$|\sum_{j=1}^{m} \beta_{j}{\bf w_{j}}|=\sqrt{1-q}$,
the right hand side of (\ref{cheq}) is at most
$2\sqrt{q(1-q)}\le 1$ with equality only if $q=1-q={1\over 2}$. This implies the first part 
of the lemma. From the above it also follows that 
the right hand side of Equation (\ref{cheq}) is
one iff $\sum_{i=1}^{n} \alpha_{i}{\bf v_{i}}
= \sum_{j=1}^{m} \beta_{j} {\bf w_{j}}$.
\end{proof}

We also show that the eigenvalues of $M$ are 
distributed symmetrically to zero.
The following observation will prove useful in many contexts:

\begin{lemma}\label{componentwise}
If $(a,b)M = (a',b')$ for some $a,a'\in \C^{[n|}$,
$b,b'\in \C^{|m]}$, then 
\[
(a,0)M=(0,b'), \;\;\;\;\;\;\; (b,0)M=(a',0).
\]
\end{lemma}

\begin{proof}
Since the diagonal
blocks of $M$ are zero, $(a,0)M$ is of the form $(0,b'')$ and 
$(0,b)M$ is of the form $(a'',0)$, which implies $(a,b)M = (a'',b'')$. 
But $(a,b)M = (a',b')$, which gives $a''=a'$, $b''=b'$, as needed.
\end{proof}

\begin{lemma}
If $(a,b)$, $a\in \C^{[n|}$, $b\in \C^{|m]}$ is an eigenvector 
of $M$ with eigenvalue $\lambda$ then
$(a,-b)$ is an eigenvector of $M$ with eigenvalue $-\lambda$.
\end{lemma}

\begin{proof}
From $(a,b)M= (\lambda a,\lambda b)$ Lemma
\ref{componentwise} gives that 
$(a,0) M = ( 0 ,\lambda b)$ and $(0,b)M= (\lambda a,0)$. Then
$(a,-b)M= (-\lambda a,\lambda b)= -\lambda ( a,- b)$.
\end{proof}

We denote the eigenvalues of $M$ by
$1 \ge \lambda_{1} \ge \lambda_{2} \ge \ldots\ge 
\lambda_{m+n}$ and with $d_{\lambda}$ the dimension of the eigen-space 
associated with eigenvalue $\lambda$.

Let us now undertake the task of computing the 
spectrum and eigenvalues of
\[
\mu = \rf_{\A}\rf_{\B} = (2\RW - I)(2\CL - I).
\]
Instead of $\mu$ it will be slightly more convenient to analyze the operator
$\kappa = {1\over 2}(\mu - I) = 2\RW\CL -\RW - \CL$ and its action on the
busy subspace. Recall the definition of the tilde operation from
Section \ref{spectralsection}.
From Lemma \ref{imagelemma}:

\begin{eqnarray}\label{eq1}
\tilde{a} \RW & = & \tilde{a}; \\
\tilde{b} \CL & = & \tilde{b}.
\end{eqnarray}

The proof of the theorem is powered by the following relations:

\begin{eqnarray}\label{eq3}
\tilde{a} \RW & = &  ((a,0) M)^{\sim};\\\label{eq4}
\tilde{b} \CL & = &  ((0,b) M)^{\sim}.
\end{eqnarray}

\begin{proof}
We prove only Equation (\ref{eq3}), the proof of 
(\ref{eq4}) is analogous. Because of linearity
it is enough to prove  (\ref{eq3}) for the 
basis vectors. Let $1\le i\le n$ be arbitrary
and let $a = (0,\ldots,0,1,0,\ldots,0|$ be the unit vector,
with a $1$ in the $i^{\rm th}$ position. Then $\tilde{a} = 
{\bf v}_{i}$. 
We have:
\[
\tilde{a}\CL  = \sum_{j=1}^{m} {\bf v}_{i} {\bf w}_{j}^{\ast}{\bf w}_{j}
=  \sum_{j=1}^{m} \langle {\bf v}_{i},\; {\bf w}_{j} \rangle
{\bf w}_{j} =  \sum_{j=1}^{m} M [{\bf v}_{i}, {\bf w}_{j}]{\bf w}_{j}.
\] 
Above  $M[{\bf v}_{i}, {\bf w}_{j}]$ means the entry of $M$
indexed by the row associated with ${\bf v}_{i}$
and by the column associated with ${\bf w}_{j}$.
Thus $\tilde{a} \CL = \tilde{b}$, where
\[
b = | M [{\bf v}_{i}, {\bf w}_{1}],\ldots, M [{\bf v}_{i}, {\bf w}_{m}]).
\]
But this $b$ is exactly $(a,0)M$. (More precisely, $(a,0)M= (0,b)$.)
\end{proof}

First we look at the action of $\kappa$ on $\A\cap \B$:

\begin{lemma}\label{intersectlemma}
We characterize $\A\cap\B$ different ways:
\begin{enumerate} 
\item A vector is in $\A\cap\B$ if and only if it can be
written both as $\tilde{a}$ and 
$\tilde{b}$.
\item Every vector in $\A\cap\B$ is
an eigenvector of $\kappa$ with eigenvalue 0. (Or, equivalently,
an eigenvector of $\mu$ with eigenvalue 1.)  
\item $\tilde{a} = \tilde{b}$
iff $(a,b)M = (a,b)$, i.e. $(a,b)$ is an eigenvector 
of $M$ with eigenvalue 1.
\end{enumerate}
\end{lemma}

\begin{proof} 1. is true by definition. For 2.
assume that $\tilde{a} = \tilde{b}$.
Then $\tilde{b} \RW = \tilde{a} \RW = \tilde{a}$ and 
$\tilde{a} \CL = \tilde{b} \CL = \tilde{b}$. 
Hence

\[
2\tilde{a}\RW\CL - \tilde{a}\RW - \tilde{a}\CL =
2\tilde{a}\CL - \tilde{a} - \tilde{b} = 2\tilde{b} - \tilde{a} - \tilde{b}
= 0,
\]

\noindent
since $\tilde{a} = \tilde{b}$. Next we prove item 3. If 
$\tilde{a} = \tilde{b}$ then $\tilde{b} \RW = \tilde{a} \RW = \tilde{a}$,
so by (\ref{eq3}) also $(0,b)M = (a,0)$. Similarly,
$\tilde{a} \CL = \tilde{b}$, so by (\ref{eq4})
$(a,0)M = (0,b)$. Therefore $(a,b)M = (a,b)$.
Conversely, if 
$(a,b) = (\alpha_{1},\ldots,\alpha_{n},\beta_{1},\ldots,\beta_{m})$ 
is an eigenvector of $M$ with 
eigenvalue 1, then by Lemma \ref{spectrumlemma}
$\tilde{a} = \tilde{b}$.
\end{proof}

Next we shall create eigenvectors for $\kappa$
from the eigenvectors for $M$ with eigenvalue less than 1.
Let  $(a,b)$, $a\in \C^{[n|}$, $b\in \C^{|m]}$ be an eigenvector 
of $M$ with eigenvalue $\lambda\neq 0$, i.e.
\[ 
(a,b)M = (\lambda a,\lambda b). 
\]

Then Lemma \ref{componentwise} gives:

\begin{eqnarray}
(a,0)M = (0,\lambda b); \\
(0,b)M = (\lambda a,0). 
\end{eqnarray}

Combining these with (\ref{eq3}) and (\ref{eq4}) we get:

\begin{eqnarray}\label{n3}
\tilde{a} R & = &
\lambda \tilde{b}; \\\label{n4}
\tilde{b} C & = & 
\lambda \tilde{a}; 
\end{eqnarray}

We would like to find a $\beta$ such that 
$v = \tilde{a} + \beta\tilde{b}$ is an eigenvector of $\kappa$.
We use (\ref{n3})-(\ref{n4}) to compute $v \kappa$:

\begin{eqnarray}
v \kappa = (\tilde{a} + \beta\tilde{b}) (2\RW\CL - \RW - \CL) & = & \\
2 \tilde{a} \RW\CL - \tilde{a} \RW - \tilde{a} \CL  + 2\beta \tilde{b} \RW\CL
-\beta \tilde{b} \RW - \beta \tilde{b} \CL & = & \\
2\lambda\tilde{b} - \tilde{a} - \lambda \tilde{b} 
+2\beta\lambda^{2}\tilde{b} -\beta\lambda \tilde{a}
- \beta \tilde{b} & = & \\
( -1 - \beta \lambda ) \tilde{a} + 
(\lambda + 2\beta \lambda^{2} -\beta) \tilde{b}.
\end{eqnarray}

We conclude that as long as
\begin{equation}\label{betaeq}
\lambda + 2\beta \lambda^{2} -\beta = \beta ( -1 - \beta \lambda ),
\end{equation}
$v$ is an eigenvector of $\kappa$ with eigenvalue
$ -1 - \beta \lambda $. Let us express $\beta$ from (\ref{betaeq}):
\begin{equation}\label{quadratic}
\beta^{2}\lambda + 2\beta \lambda^{2} + \lambda = 0.
\end{equation}

Solving the equation gives $\beta = - \lambda \pm \sqrt{\lambda^{2} - 1}$.
Considering that $\mu = 2\kappa + I$ we can now write down the
eigenvectors and eigenvalues of $\kappa$ and $\mu = 2\kappa +I$ 
that we obtain from eigenvectors and eigenvalues of $M$.
We summarize the formulas in the following two tables.
The first table refers to the case of $\lambda\in (0,1)$:

\medskip

\begin{center}
\begin{tabular}{l|c|c|}
 & Eigenvector & Eigenvalue \\\hline\hline
$M$ & $(a,b)$ & $\lambda$ \\
 & $(a,-b)$ & $-\lambda$ \\\hline
$\kappa$ & $\tilde{a} - \lambda \;\tilde{b} + \, i\, 
\sqrt{1- \lambda^{2}}\;\tilde{b}$ &
$\lambda^{2}-1  - \, i\, \lambda \sqrt{1- \lambda^{2}}$ \\
  & $\tilde{a} - \lambda \; \tilde{b} - 
\, i\, \sqrt{1- \lambda^{2}}\;\tilde{b}$ &
$\lambda^{2}-1  + \, i\, \lambda \sqrt{1- \lambda^{2}}$ \\\hline
$\mu$ & $\tilde{a} - \lambda \; \tilde{b} + \, i\, 
\sqrt{1- \lambda^{2}}\; \tilde{b}$ &
$2 \lambda^{2}-1  - 2 \, i\, \lambda \sqrt{1- \lambda^{2}}$ \\
  & $\tilde{a} - \lambda \; \tilde{b} - \, i\, \sqrt{1- \lambda^{2}}
\; \tilde{b}$ &
$2 \lambda^{2}-1  + 2 \, i\, \lambda \sqrt{1- \lambda^{2}}$
\end{tabular}
\end{center}
\smallskip

The second table refers to the case when $\lambda=0$. In this case
$(a,b)M=(0,0)$ implies that $(a,0)M=(0,0)$ and  $(0,b)M=(0,0)$.
Therefore the zero subspace of $M$ decomposes into 
the direct sum of the (possibly 0-dimensional)
subspaces $\{(a,0)\mid (a,0)M=(0,0)\}$ and 
$\{(0,b)\mid (0,b)M=(0,0)\}$.
Since Equation (\ref{quadratic}) in this special case holds with every $\beta$,
we obtain:

\medskip

\begin{center}
\begin{tabular}{l|c|c|}
 & Eigenvector & Eigenvalue \\\hline\hline
$M$ & $(a,0)$ & 0 \\
 & $(0,b)$ & 0 \\\hline
$\kappa$ & $\tilde{a}$ & $-1$ \\
 & $\tilde{b}$ & $-1$ \\\hline
$\mu$ & $\tilde{a}$ & $-1$ \\
 & $\tilde{b}$ & $-1$
\end{tabular}
\end{center}

\smallskip

We are left to show that we have found all $n+m-\dim (\A\cap\B)$ 
orthogonal eigenvectors of the busy subspace.
In the above tables we lined up
the eigenvectors of $\kappa$ with the eigenvectors
of $M$  from which they originate to suggest
a one-one correspondence. Observe that among the eigenvalues the 
correspondence is established by
\[
(\mbox{$M$-side})
\;\;\;\;\;\;\;\;\;\;\;
\pm\, \lambda\;\;\;\;\; \longleftrightarrow \;\;\;\;\;
\lambda^{2}-1\, \pm \, i \,\lambda \sqrt{1- \lambda^{2}}
\;\;\;\;\;\;\;\;\;\;\;
(\mbox{$\kappa$-side})
\]

The numbers seem to match,
since $M$ has  $n+m - d_{1} - d_{-1}= n+ m - 2d_{1}$ eigenvectors 
with eigenvalues in the range $(-1,1)$. These correspond to
the the same number of 
eigenvectors for $\kappa$ in the busy subspace
with non-zero eigenvalues. In addition, the busy subspace contains
$d_{1}$ independent eigenvectors with eigenvalue zero.
We get a total of $n+m-d_{1}$ eigenvectors. By $\dim(\A+\B)=
\dim\A +\dim\B-\dim(\A\cap\B)$ and 
and Lemma \ref{intersectlemma} the dimension of the busy subspace is also 
$n+m-d_{1}$.
We cannot walk away from the task, however, of showing that no
dependencies occur among the eigenvectors we
constructed. Since eigen-spaces associated
with different eigenvalues are orthogonal, is sufficient to show that 

\begin{lemma}\label{dimen}
Let $|\lambda|<1$. Then the dimensions of the 
eigen-spaces of $\kappa$
associated with eigenvalues
$\lambda^{2}-1\, \pm \, i \,\lambda \sqrt{1- \lambda^{2}}$
are $d_{\lambda}$ (each). Also,
the dimension of the 
$-1$-eigen-space of $\kappa$ is $d_{0}$.
\end{lemma}

\begin{proof}
Let $\lambda\in (0,1)$ and
\[
S_{\lambda} = \{(a,b)\mid a\in \C^{[n|},\; b\in\C^{|m]},\; 
(a,b)M = \lambda (a,b)\}.
\]

Let $\tau_{+}$ and $\tau_{-}$ be the operators from $S_{\lambda}$ to
$\C^{[n|\times |m]}$
defined by 
\begin{eqnarray*}
\tau_{+} : \;\;\; (a,b) \;\;\;\; \rightarrow \;\;\;\;
\tilde{a} - \lambda \tilde{b} + i \sqrt{1- \lambda^{2}}\;\tilde{b}. \\
\tau_{-} : \;\;\; (a,b) \;\;\;\; \rightarrow \;\;\;\;
\tilde{a} - \lambda \tilde{b} - i \sqrt{1- \lambda^{2}}\;\tilde{b}.
\end{eqnarray*}

We need to show that the images of $\tau_{+}$ and $\tau_{-}$ 
have dimension $d_{\lambda} = \dim S_{\lambda}$. 
We prove that the kernel of both $\tau_{+}$ and $\tau_{-}$
are trivial. We show this only for $\tau_{+}$, since
the proof for $\tau_{-}$ goes in the same way.
Let us assume, contrary to the lemma, that for some $a\in\C^{[n|}$,
$b\in\C^{|m]}$, $(a,b)\in S_{\lambda}$, $(a,b)\neq (0,0)$ we have 
$(a,b) \tau_{+} = \tilde{a} - \left(\lambda - i \sqrt{1- \lambda^{2}}
\right)\;\tilde{b}=0.$ In fact it is enough to show that
the assumption implies $a=0$ or $b=0$, since both imply the other.
By Lemma \ref{intersectlemma}
$(a,\left(\lambda - i \sqrt{1- \lambda^{2}}
\right)b)$ is an eigenvector of $M$ with eigenvalue 1, which implies
$(a,0)M = (0,\left(\lambda - i \sqrt{1- \lambda^{2}}
\right)b)$. Since $(a,b)\in S_{\lambda}$, we also have
$(a,0)M = (0,\lambda b)$. Hence
\[
\lambda b = \left(\lambda - i \sqrt{1- \lambda^{2}}
\right)b,
\]
which, since $|\lambda| < 1$ can happen only if 
$b=0$, a contradiction. Note that the proof works for 
the $\lambda=0$ case too.
\end{proof}

\section{Norms and Inner Products}\label{normsection}

In this section we show how to compute norms of vectors
in $\A+\B$, and in particular we compute the norms of the 
eigenvectors we obtained in the previous section.
We show that
$\tau_{+}$ and $\tau_{-}$ 
are scalar product preserving up to a constant scaling factor,
and determine this constant.
This gives an alternative proof to Lemma \ref{dimen}.

\begin{lemma}\label{scalarlemma}
For any $a,a'\in  \C^{[n|}$, $b, b'\in  \C^{|m]}$ it holds that
\begin{eqnarray}\label{scalar1}
\langle \tilde{a},\tilde{a'}\rangle & = & \langle a,\; a'\rangle ; 
\\\label{scalar11}
\langle \tilde{b},\tilde{b'}\rangle & = & \langle b,\; b'\rangle ; 
\\\label{scalar2}
\langle \tilde{a},\tilde{b}\rangle  & = & (a,0)M(0,b)^{\ast}.
\end{eqnarray}

\noindent Furthermore, if $(a,b), (a',b')\in S_{\lambda}$:

\begin{eqnarray}\label{abrel1}
\langle a,\, a'\rangle & = & \langle b,\, b'\rangle \\\label{abrel2}
\langle \tilde{a},\,\tilde{b'}\rangle 
& = & \lambda \langle a,\, a'\rangle
\end{eqnarray}
\end{lemma}

\begin{proof}
Indeed, for $a=(\alpha_{i}|$, $a'=(\alpha'_{i}|$, 
$b=|\beta_{i})$, $b'=|\beta'_{i})$
\begin{eqnarray*}
\langle \tilde{a},\tilde{a'}\rangle
& = & \sum_{1\le i\le n}\alpha_{i} {\bf v}_{i} {\bf v}_{i}^{\ast} 
\overline{\alpha'_{i}} =
\sum_{1\le i\le n} \alpha_{i}\overline{\alpha'_{i}}
= \langle a\; a' \rangle ; \\
\langle \tilde{b},\tilde{b'}\rangle
& = & \sum_{1\le j\le m}\beta_{j}{\bf w}_{j} {\bf w}_{j}^{\ast} 
\overline{\beta'_{j}} =
\sum_{1\le i\le n} \beta_{j}\overline{\beta'_{j}}
= \langle b\; b' \rangle ; \\
\langle \tilde{a},\tilde{b}\rangle & = & 
\sum_{1\le i\le n} \sum_{1\le i\le m}
\alpha_{i} \;{\bf v}_{i} {\bf w}_{j}^{\ast} \;\overline{\beta_{j}} =
\alpha_{i} \;\langle {\bf v}_{i}, {\bf w}_{j}\rangle \;\overline{\beta_{j}} 
= (a,0)M(0,b)^{\ast}.
\end{eqnarray*}

Consider now any $(a,b), (a',b')\in S_{\lambda}$. We have
$(a,0)M(0,b')^{\ast}= (0,\lambda b) (0,b')^{\ast}=  \lambda \langle b,\; 
b'\rangle.$
But also, $(a,0)M(0,b')^{\ast}= (a,0) ((0,b')M^{\ast})^{\ast}
= (a,0) ((0,b')M)^{\ast} = (a,0)(\lambda a',0)^{\ast} = 
\lambda \langle a, \; a'\rangle$. By the virtue of
(\ref{scalar2}) the above shows not only (\ref{abrel1}) but also 
(\ref{abrel2}).
\end{proof}

Using Lemma \ref{scalarlemma}:
\begin{eqnarray}\nonumber
\langle \tilde{a} - \lambda \tilde{b} + i \sqrt{1- \lambda^{2}}\;\tilde{b},
\tilde{a'} - \lambda \tilde{b'} + i \sqrt{1- \lambda^{2}}\;\tilde{b'} \rangle
 & = & \\\label{normexpression}
\langle \tilde{a},\tilde{a'} \rangle +
\langle \tilde{b},\tilde{b'} \rangle 
-\lambda \langle \tilde{b},\tilde{a'} \rangle 
-\lambda \langle \tilde{a},\tilde{b'} \rangle.
\end{eqnarray}
From Lemma \ref{scalarlemma} 
$\langle \tilde{b},\tilde{a'} \rangle =
\overline{\langle \tilde{a'},\tilde{b} \rangle } =
\overline{\lambda \langle \tilde{a'},\tilde{a} \rangle }
= \lambda \langle \tilde{a},\tilde{a'} \rangle$.
By introducing 
$\gamma = \langle \tilde{a},\tilde{a'} \rangle$ we can write 
Expression (\ref{normexpression}) as $\gamma(2-2\lambda^{2})$.

On the other hand 
\[
\langle (a,b),(a',b')\rangle = \langle a,a'\rangle
+ \langle b,b'\rangle = 2\gamma.
\]

We conclude that the scaling factor is $1-\gamma^{2}$ i.e.
for every $v,w\in S_{\lambda}$:

\[
\langle \tau_{+}(v), \tau_{+}(w) \rangle = (1-\lambda^{2}) \langle v, w\rangle.
\]

Similarly we obtain that for every $v,w\in S_{\lambda}$:

\[
\langle \tau_{-}(v), \tau_{-}(w) \rangle = 
(1-\lambda^{2}) \langle v, w\rangle.
\]

In particular:

\begin{theorem}\label{normth}
The eigenvectors of $\mu$ in Theorem \ref{mainth} have the norm 
\medskip

\begin{tabular}{lp{4.2in}}
$\sqrt{1-\lambda^{2}}$, & if $|(a,b)| = 1$ and 
$0\le  \lambda < 1$; \\
$1$, & if $\tilde{a}\in \A\cap \B$ and  $|a|=1$.
\end{tabular}
\end{theorem}

\section{An Example}\label{illustration}

In this section we give an example to the use of Theorem
\ref{mainth} for a Markov chain
associated with Grover's algorithm. Here we present
the ``concise version'' of the chain which we call the {\em Grover Chain}.
The Grover chain has two states: $marked$ and $unmarked$.
The ``full version,'' where different
items correspond to different states has similar analysis.
(We do not give a precise mathematical justification 
of the fact that
clumping together all marked items and all unmarked items
in the way we do gives formulas similar to 
those coming from the analysis of 
Algorithm FindMarked($K$) for the chain
$P = {1\over n}E$, where $E$ is the all one matrix.
Our example is interesting on its own right even 
without this connection.)

\begin{figure}[t]\label{groverfig}
\begin{center}
\begin{tabular}{ccc}
\epsfig{file=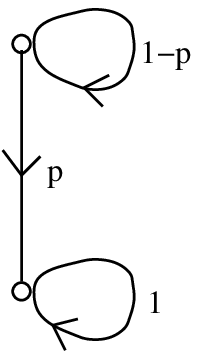} & \hspace{0.6in} & \epsfig{file=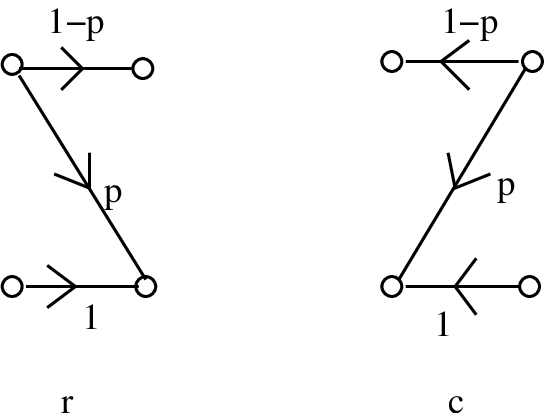}\\ 
Grover chain  &  & Bipartite version \\
\end{tabular}
\caption{The Markov chain associated with Grover's algorithm, where
items are marked with probability $p$}
\end{center}
\end{figure}

Assume that the probability that an item is marked is $p$.
The chain corresponds to the classical (non-quantum) algorithm, where
at each step we move to a random item, but when we find a marked item
we never move away from it. 
The transition of this chain takes
an unmarked item to an unmarked item with probability $1-p$ and
to a marked item with probability $p$.
On the other hand marked items alway go into marked items with probability 1.
Figure \ref{groverfig} shows the Markov chain and its associated 
bipartite maps, $\rw$ and $\cl$.
The pair describing the walk is $(\rw,\cl)$, where
\[
\rw = \cl =
\left(
\begin{array}{cc}
1-p & p \\
0 & 1 \\
\end{array}
\right).
\]
The half discriminant matrix of the walk is 
\[
D = \left(
\begin{array}{cc}
1-p & 0 \\
0 & 1 \\
\end{array}\right)
\]
With eigenvectors $a_{1} = (1,0)$ and $a_{2} = (0,1)$,  and
eigenvalues $1-p$ and $1$ respectively.
Let $b_{1} = (1,0)$, $b_{2} = (0,1)$.
Then the eigenvectors of the discriminant matrix 
$M = \left(
\begin{array}{cc}
0 & D \\
D^{T} & 0
\end{array}\right)$
are
\medskip

\begin{tabular}{lllll}
$(a_{1},b_{1})$ & with eigenvalue $1-p$; & \hspace{0.4in} &
$(a_{1},-b_{1})$ &  with eigenvalue $p-1$; \\
$(a_{2},b_{2})$ &  with eigenvalue $1$;  &  &
$(a_{2},-b_{2})$ &  with eigenvalue $-1$. \\
\end{tabular}
\medskip

Let $\nu$ be the quantized version of this bipartite chain.
By Theorem \ref{mainth} the busy subspace of $\nu$
has eigenvectors: $\tilde{a_{1}}-\tilde{b_{1}} 
+p\;\tilde{b_{1}} \pm i\sqrt{2p-p^{2}}\;\tilde{b_{1}}$
and $\tilde{a_{2}} $. The latter coincides with $\tilde{b_{2}}$.
Representing these as rectangular vectors:
\begin{eqnarray*}
& v_{1} & = 
\begin{array}{||cc||}
\sqrt{1-p}\;( p + i\sqrt{2p-p^{2}}) & \sqrt{p} \\
\sqrt{p}\;(-1 + p + i\sqrt{2p-p^{2}}) & 0
\end{array}\;; \hspace{0.2in} 
v_{2} = 
\begin{array}{||cc||}
\sqrt{1-p}\;( p - i\sqrt{2p-p^{2}}) & \sqrt{p} \\
\sqrt{p}\;(-1+ p - i\sqrt{2p-p^{2}}) & 0
\end{array}\;; \\
 & v_{3} & = 
\begin{array}{||cc||}
0 & 0 \\
0 & 1 \\
\end{array}\;. 
\end{eqnarray*}
One can compute that $|v_{1}|^{2} = |v_{2}|^{2} = 4p-2p^2$.
This is consistent with Theorem \ref{normth}, since the
squared norm of $(a_{1},b_{1})$ is 2, and 
Theorem \ref{normth} implies that the squared norms
of $v_{1}$ and $v_{3}$ can be obtained from the norm of $(a_{1},b_{1})$
by scaling it with the factor $1-(1-p)^{2}$. Theorem \ref{mainth}
gives that the eigenvalues associated with $v_{1}$, $v_{2}$ and $v_{3}$
are $1-4p+2p^{2}-2i(1-p)\sqrt{2p-p^{2}}$, 
$1-4p+2p^{2}+2i(1-p)\sqrt{2p-p^{2}}$ and $1$.
What makes Grover's algorithm work is that 
$1-4p+2p^{2}\pm 2i(1-p)\sqrt{2p-p^{2}} = e^{\pm i\theta}$, where
$\theta \in \Theta(\sqrt{p})$. Our version of Grover's algorithm
(Section \ref{memorysec}) uses the initial state
\[
u = 
\begin{array}{||cc||}
1-p & \sqrt{p(1-p)} \\
\sqrt{(1-p)p} & p
\end{array}\;.
\]
This is the state we can easily produce (in the query model
without cost, in the circuit model with small cost).
Notice that $u$ has a large component in the space 
spanned by $v_{1}$ and $v_{2}$.
Define
\[
u' = {-i\over \sqrt{2}\;|v_{1}|} v_{1} + 
{i\over \sqrt{2}\;|v_{2}|} v_{2} =
\begin{array}{||cc||}
\sqrt{1-p} & 0 \\
\sqrt{p} & 0 \\
\end{array}\;.
\]
Then
\[
\left\langle u\;,\;u' \right\rangle = \sqrt{1-p}.
\]
This again, should not surprise us because of Equation
(\ref{componenteq}). We have
\[
u' \nu^{K} = {-i e^{i\theta K} \over \sqrt{2}\;|v_{1}|} v_{1}
+ {i e^{-i\theta K} \over \sqrt{2}\;|v_{2}|} v_{2},
\]
and our analysis of Algorithm FindMarked($K$) can go on
like in the previous section. The goal of the
present article is exactly to show how to shortcut much 
of the calculations we have made in this section.

\end{document}